\documentclass{elsart}
\usepackage{amssymb}
\usepackage{epsfig}

\newcommand\eps{\epsilon}
\newcommand\beq{\begin{equation}}
\newcommand\eeq{\end{equation}}

\begin{document}
\begin{frontmatter}

\title{Fluctuations of energy injection rate in a shear flow}

\author{J\"org Schumacher and Bruno Eckhardt}
\address{Fachbereich Physik,
Philipps-Universit\"at, D-35032 Marburg, Germany}

\begin{abstract}
We study the instantaneous and local energy injection in a turbulent shear
flow driven by volume forces. The energy injection can be both positive
and negative. Extremal events are related to coherent streaks.
The probability distribution is asymmetric, deviates slightly from a
Gaussian shape and depends on the position in shear direction. The
probabilities for positive and negative injection are exponentially
related, but the prefactor in the exponent varies across the shear layer. 
\end{abstract}
\end{frontmatter}

\section{Introduction}
Turbulent flows are a characteristic example for a dissipative
macroscopic system that is driven steadily far from equilibrium. Energy
of motion is dissipated continuously due to friction between the fluid parcels
and has to be fed in by some kind of stirring or a large scale 
shear gradient in order to
sustain the turbulent state \cite{Frisch}.
Although this picture is as old as
turbulence research itself, not much is known about the interplay between the rates
of energy dissipation and energy input. 
Several recent experimental efforts have focussed on studies of the
energy input rate into turbulent systems:  Ciliberto and Laroche \cite{Ciliberto98}
extracted the buoyant energy input in turbulent Rayleigh-B\'{e}nard convection,
Goldburg {\it et al.}  \cite{Goldburg} measured the power transfered to a liquid
crystal in a chaotic regime above the electrohydrodynamic instability,
Pinton {\it et al.} and Cadot {\it et al.} studied the injection rate statistics
with von Karman swirling flow experiments
\cite{Pinton98,Pinton99,Pinton2002,Cadot02}.

On the level of a macroscopic description energy dissipation in a 
turbulent flow is always positive, reflecting the irreversibiliy of the
dynamics through a direct relation between dissipation and entropy production.
While in the mean energy dissipation and energy uptake are equal, the fluctuations
of energy uptake are much less constraint and can take both signs.

Fluctuations of the energy dissipation have recently attracted attention
in connection with the observation that in the case
of reversible non-equilibrium systems both signs for the entropy
production are possible, but that the entropy reduction is exponentially
suppressed compared to entropy production \cite{Evans93,GC95}
(for a comprehensive review see \cite{Vollmer}). For hydrodynamic systems
these ideas do not strictly apply, since the 
Navier-Stokes equation is not reversible. But the study
of Farago \cite{Farago} shows that even in the absence of that symmetry, as e.g. for a
Brownian particle, interesting relations among fluctuations can arise.

It is possible to modify the
macroscopic equations, as in constrained Euler ensembles \cite{She93} or in
isokinetic Navier-Stokes ensembles \cite{Rondoni,Gallavotti02}, in order to 
arrive at a reversible macroscopic dynamics.
However, we will in the following adopt the 
traditional Navier-Stokes equation as our 
starting point, take the energy uptake as observable, as in the
experiments mentioned above, and embark on a numerical study of 
the statistics of local energy input in a turbulent shear flow 
and especially on the ratios of the probability density functions (PDF) for 
negative and positive energy injection. We will also study the relation between
extremal injection events and coherent structures.
The results presented here connect to and complement
studies of volume averaged fluctuation statistics in
hydrodynamic systems, including observations on thermal convection
\cite{Ciliberto98}, swirling flows \cite{Pinton99,Cadot02}, 
or GOY shell models \cite{Aumaitre01}.

The specific system we study here is a hydrodynamic shear flow driven
by volume forces. The system is a macroscopic version of the microscopic
molecular dynamics simulations of shear flows in Evans {\it et al.}
\cite{Evans93}. The main
differences between the microscopic and the macroscopic model are
irreversibility and the absence of a thermostat in the latter. 
The flow is incompressible and a statistically stationary
turbulent state is sustained by driving with a steady volume
force ${\bf F}=F_x(y){\bf e}_x$.
The Navier-Stokes equation then reads
\begin{eqnarray}
\label{nseq}
\frac{\partial{\bf u}}{\partial t}+({\bf u}\cdot{\bf \nabla}){\bf u}
&=&-{\bf \nabla} p+ \nu {\bf \nabla}^2{\bf u}+{\bf F}\;,\\
\label{ceq}
{\bf \nabla}\cdot{\bf u}&=&0\;.
\end{eqnarray}
${\bf u}({\bf x},t)$ is the velocity field, 
$p({\bf x},t)$ the kinematic pressure, and
$\nu$ the {constant} kinematic viscosity.
The equations of motion are solved by means of a pseudospectral
method in a volume $V$ with periodic boundary conditions in $x$ (streamwise) 
and $z$ (spanwise) directions and with free-slip boundary conditions in shear
direction $y$ (for more details see \cite{Schu01}).  
The rectangular slab has an aspect ratio of $2\pi:1:2\pi$. 
The spectral resolution is $N_x\times N_y\times N_z=256\times 65\times 256$ 
for all runs and the number of degrees of freedom (or independent Fourier modes) 
is about $2.5\cdot 10^6$. With $U_0$ the amplitude of the laminar shear 
profile, ${\bf u}_0=U_0 \cos(\pi y/L_y) {\bf e}_x$, and $L_y$ the width of the slab,
we can define a Reynolds number $Re=U_0 L_y/\nu$;
for our simulations it takes values between 500 and 6000. 

\section{Energy balance}
We want to study the changes in the total kinetic energy density of the fluid,
$e({\bf x},t) = {\bf u}({\bf x},t)^2/2\,$.
Its evolution is governed by
\beq
\partial_t e +
\partial_j(u_j e - \nu u_i \partial_j u_i + u_i p \delta_{ij})
= u_i F_i - \nu (\partial_i u_j)^2 \,.
\label{energyeq}
\eeq
For the energy balance of the volume averaged kinetic energy,
$E(t)= \langle {\bf u}^2\rangle_V/2 $
the second term on the left hand side of (\ref{energyeq}) does not contribute
as it contains
a total divergence of a current whose surface integral vanishes for the given
boundary conditions.
The balance for the volume averaged kinetic energy thus reads
\begin{eqnarray}
\label{energy}
\frac{\partial E}{\partial t}=\frac{1}{V}\int_{V} u_x F_x
\,\mbox{d}V-\frac{\nu}{V}\int_{V} ({\bf\nabla u})^2
\,\mbox{d}V=\epsilon_{in}(t)-\epsilon(t)\,.
\end{eqnarray}
The energy dissipation rate $\epsilon(t)$, and its local
version $\epsilon({\bf x},t)=(\nu/2)(\partial_i u_j+\partial_j u_i)^2$, are 
positive semi-definite (though zero values never occur in practice). Thus,
energy is steadily taken out of the system. In contrast to thermostatted systems
\cite{Evans93,Rondoni,Gallavotti02}, 
the energy injection rate $\epsilon_{in}(t)$ is not synchronized to
$\epsilon(t)$, the system is able to store energy for intermediate time intervals
and the kinetic energy content, $E(t)$, fluctuates in time.

For the volume forces studied here the local energy injection rate becomes
\begin{equation}
\epsilon_{in}({\bf x},t)= u_x({\bf x},t) F_x(y)=
\frac{\nu \pi^2 U_0}{L_y^2} \cos(\pi y/L_y) u_x({\bf x},t)\,,
\end{equation}
where $u_x({\bf x},t)$ is the streamwise velocity component.
Energy that is given back to the source corresponds to negative values
of $\epsilon_{in}$.
\begin{figure}
\begin{center}
\epsfig{file=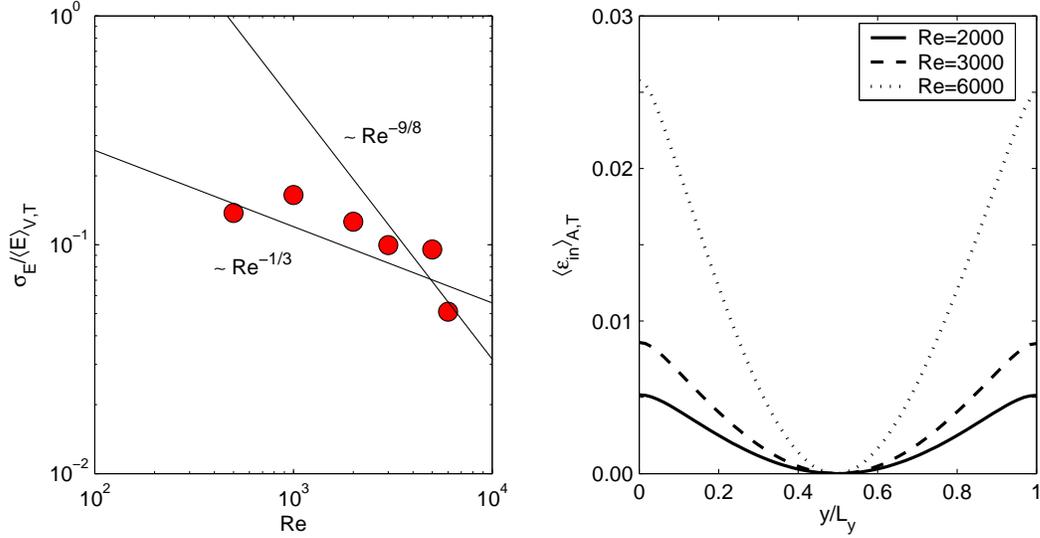, width=14cm}
\end{center}
\caption[]{Left:  Ratio of standard deviation of the kinetic energy
fluctuations, $\sigma_E$, and the time average, $\langle E
\rangle_T=\langle {\bf u}^2\rangle_{V,T}/2$, versus Reynolds number, $Re$.  
Solid
lines indicate the power laws with the corresponding exponents. Right: Mean
energy injection rate averaged in planes $A=[0,L_x]\times [0,L_z]$ at fixed
$y$ and in time, $\langle\epsilon_{in}\rangle_{A,T}$, as a function of the
shear direction $y$. Reynolds numbers are indicated in the legend. The 
particular kind of steady forcing gives rise to a $y$-dependent profile for the
mean energy injection rate.}
\label{fig2}
\end{figure}
 
The variances of the fluctuations in the kinetic energy are shown for different
values of the Reynolds number in Fig.~\ref{fig2}.  In addition to the volume
average $\langle\cdot\rangle_V$ we will also use averages over planes at fixed
$y$ and over time or combinations of spatial and temporal averages which are
thought as statistical ensemble averages.  Those averages will be indicated by
$\langle\cdot\rangle_A$ and $\langle\cdot\rangle_T$, respectively.  The
fluctuations of volume averaged kinetic energy (left panel of Fig.~\ref{fig2})
decrease with increasing Reynolds number.  This is an indication that the
increasing number of degrees of freedom that become dynamically active with
increasing Reynolds number tend to average out.  In quantitative terms the
number of active Fourier modes outside the viscous subrange (i.e.  with wave
lengths exceeding the Kolmorgorov scale $\eta_K=(\nu^3/\epsilon)^{1/4}$) is
roughly $N\sim (L/\eta_K)^3\sim Re^{9/4}$ \cite{Landau}, so that, assuming a law
of large numbers, we might expect $\sigma_E\sim N^{-1/2}\sim Re^{-9/8}$.
Experiments in a closed swirling turbulent flow indicated a slope $\sim
Re^{-1/3}$ \cite{Pinton99}.  For the Reynolds number range accessible in our
simulations no definite conclusion on either scaling can be drawn.

For the applied volume forcing the flow is not homogeneous in the shear
direction, as indicated by the variations of average energy uptake with position
in shear direction (right panel of Fig.~\ref{fig2}).  It is homogeneous in the
downstream and the spanwise direction, so that for all statistical measures we
can collect in one ensemble all points in the $x$-$z$-planes for fixed shear
coordinate $y$.

\begin{figure}
\begin{center}
\epsfig{file=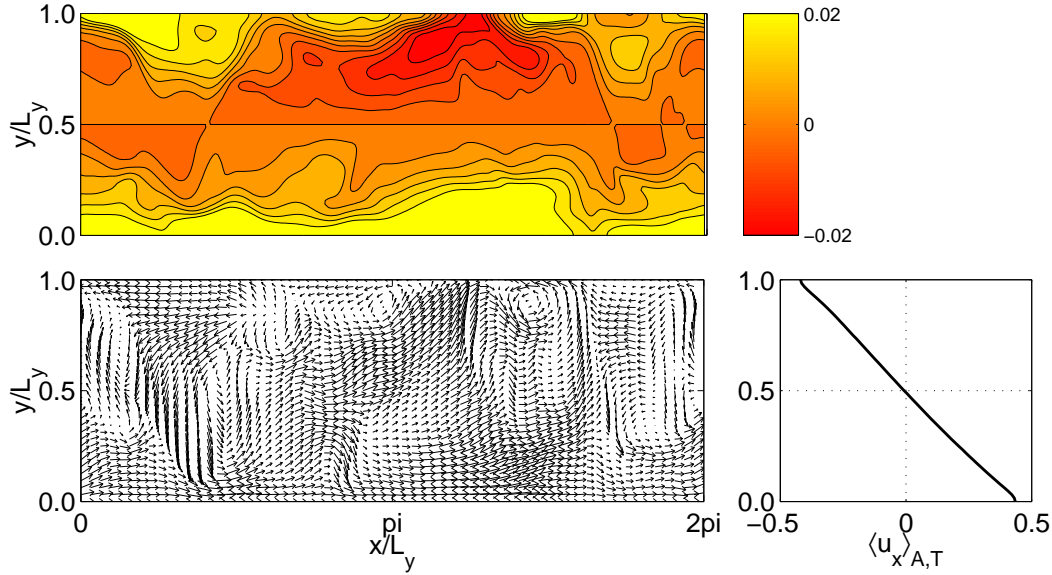, width=14cm}
\end{center}
\caption[]{Snapshot of the velocity field and the corresponding energy
injection rate field at fixed position $z_0/L_y=\pi$ for the run with
$Re=6000$.  The upper panel shows $\epsilon_{in}$ and the lower left panel shows
$u_x$ and $u_y$ in a vector plot. 
The mean flow profile $\langle u_x\rangle_{A,T}(y)$
is indicated in the lower right.
Clearly, the dark regions with negative
$\epsilon_{in}$ conincide with regions where  
the fluid moves coherently opposite to the mean flow.
}
\label{fig1}
\end{figure}

The higher energy content of the largest scales and the prevalence
of large coherent structures in shear flows \cite{Schu00} suggest
that the energy uptake is dominated by large scale flows.
In shear flows, the predominant structures are downstream
vortices and streamwise streaks. One such event that is connected
with a negative energy injection rate is shown in Fig.~\ref{fig1}.
We do observe fluid moving coherently into the opposite direction
to the mean flow $\langle u_x\rangle_{A,T}(y)$ which is plotted
in the lower right panel of Fig.~\ref{fig1}.

\section{Statistical analysis of the energy injection rate}

The energy input can be large, both positive and negative, but it cannot
be arbitrarily large. Its volume average, 
$\epsilon_{in}(t)$, can be related to the rms velocity fluctuations
by a Cauchy-Schwartz inequality
\begin{eqnarray}
|\langle \epsilon_{in}(t)\rangle_T|
\le \left\langle \langle u_x^2\rangle_V^{1/2} \langle F_x^2\rangle_V^{1/2}
\right\rangle_T
=\frac{\nu \pi^2 U_0}{\sqrt{2} L_y^2} \langle u_{x, rms}\rangle\,.
\label{Cauchy}
\end{eqnarray}
Fluctuations exceeding the external velocity
scale $U_0$ are extremely unlikely, so that $\epsilon_{in}$ is
bounded by $\sim U_0^2$. The local energy injection rate can be 
estimated using the maximum norm, with a similar bound $\sim U_0^2$,
but a different prefactor.
For the numerical simulations both bounds turned out to be much
too crude and much larger than both the maximal and the rms fluctuations
of the downstream velocity. These bounds would thus become effective
in the far tails of the probability distribution only.

\begin{figure}
\begin{center}
\epsfig{file=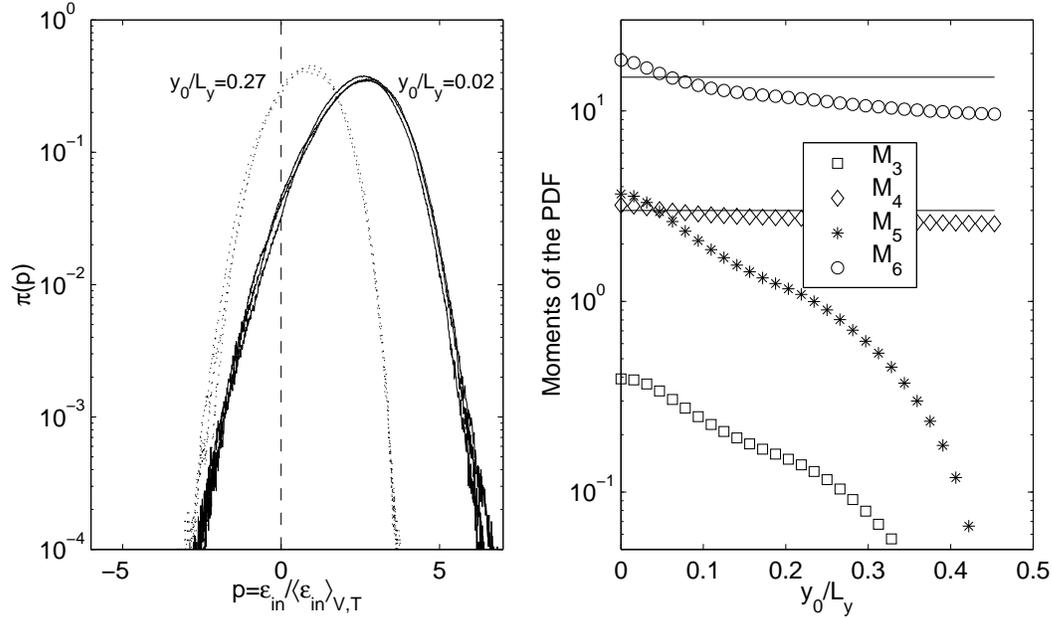, width=14cm}
\end{center}
\caption{Statistics of energy uptake for different vertical 
positions in the shear layer.
Left: Probability density function (PDF) of the energy input rate.
The statistics is based on about 100 turbulent snapshots separated in time by
1.5 large eddy turnover times; it contains about $6\times 10^8$ data points.
The PDF's are shown for three different Reynolds numbers, $Re=3000$, $5000$ and 
$6000$ and the $y_0$ values of the fixed planes are indicated in the figure.
Right: Normalized moments of $\epsilon_{in}$
versus position $y_0/L_y$ of fixed plane. Values for 
a Gaussian distribution, $M_4=3$ and $M_6=15$, are indicated by solid lines.
Reynolds number was 6000.}
\label{fig4}
\end{figure}

We focus on the fluctuations of the instantaneous, pointwise energy injection
$\eps_{in}({\bf x},t)$. We expect that any kind of spatial or temporal average
over regions and times longer than the corresponding correlation lengths
and times would push the distributions closer to Gaussian shape.
It is interesting to note that this does not seem to be the 
case in \cite{Pinton98,Pinton99}. But even there an instantaneous
and local distribution should provide a more sensitive measure of
possible deviations from a Gaussian shape.
Since our system is invariant under translation in 
downstream and spanwise direction, but not in the normal
direction, we study the distributions for planes 
parallel to the bounding surfaces separately.
The probability density functions of the energy input rate in units
of its ensemble average, $\langle\epsilon_{in}\rangle_{V,T}$, and for 
different positions between the plates are shown in Fig.~\ref{fig4}.  

The collapse of the PDF's for different Reynolds numbers, rescaled by the
volume averages, indicates a universality of the fluctuations of large 
scale injection (left panel of Fig.~\ref{fig4}), as also observed
in experiment \cite{Pinton99,Cadot02}. The PDF we find is similar to 
the one measured in \cite{Cadot02} and much closer to Gaussian shape 
than the one in \cite{Pinton99}. As a measure of deviations we determine the 
normalized and centered
moments $M_n=\langle (p-\langle p\rangle_{A,T})^n\rangle/
             \langle (p-\langle p\rangle_{A,T})^2\rangle^{n/2}$ where 
$\langle (p-\langle p\rangle_{A,T})^n\rangle
   =\int (p-\langle p\rangle_{A,T})^n \pi(p) \,\mbox{d}p$ and 
   $p=\epsilon_{in}/\langle\epsilon_{in}\rangle_{V,T}$.
In particular, the ones for $n=4$ and $n=6$
differ from the Gaussian values of $3$ and $15$ which are indicated by solid
lines in the right panel of Fig.~\ref{fig4}.
However, the PDF also changes when 
going from the free-slip side planes at $y_0/L_y=0$ towards the center 
at $y_0/L_y=1/2$: the odd moments are smaller near the center, indicating a
more symmetric PDF (see right panel of Fig.~\ref{fig4}). The PDF is closest to
Gaussian here.

\begin{figure}
\begin{center}
\epsfig{file=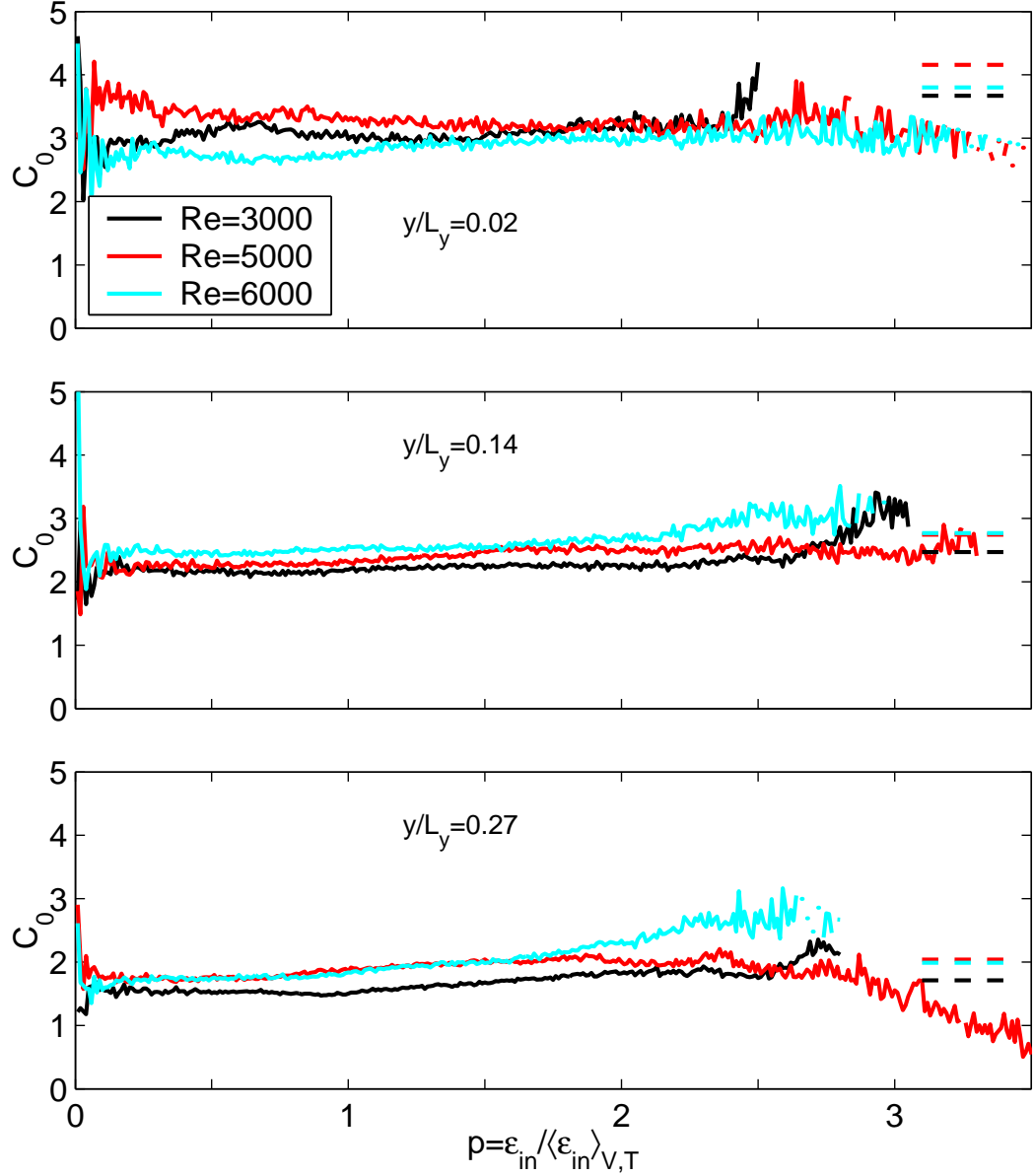, width=14cm}
\end{center}
\caption{Test of the linearity of the logarithmic ratio as suggested by 
Eq.~(\ref{c0}). The stronger oscillations for very small $p$ arise due to
the amplification of small wiggles on the PDF $\pi(p)$ there. The linear
slope is fairly independent of $Re$. Data are as in Fig.~\ref{fig4}.
The dashed lines on the far right indicate the values of $C_0$ 
obtained from Eq.~(\ref{czero}).}
\label{fig5}
\end{figure}

We next turn to the logarithm of the ratio of
the probabilities for positive and negative injection rates,
$\log [\pi(p)/\pi(-p)]$, where $p$ is again a particular value of 
the dimensionless instantaneous and local dissipation, 
$\epsilon_{in}({\bf x},t)/\langle\epsilon_{in}\rangle_{V,T}$. 
In Fig.~\ref{fig5} we show
\begin{eqnarray}
C_0(p)=\frac{1}{p} \log\left[\frac{\pi(p)}{\pi(-p)}\right]\,.
\label{c0}
\end{eqnarray}
A constant $C_0$ indicates a linear exponential relation
between the probabilities of positive and negative energy injection.
As is well known the slope $C_0$ can be related to the mean and
the variance in a Gaussian model for the fluctuations. Clearly the logarithmic
ratio of the probability density functions is then automatically linear. 
If the distribution has the form
\beq
\pi(p) = N \exp(-(p-\overline{p})^2/2\sigma^2)\,,
\eeq
with normalization factor $N$, then
\beq
C_0 = 2 \frac{\overline{p}}{\sigma^2} = 2\frac
{\langle\epsilon_{in}\rangle_{A,T}\langle\epsilon_{in}\rangle_{V,T}}
{\langle\epsilon_{in}^2\rangle_{A,T}-\langle\epsilon_{in}\rangle_{A,T}^2} \,.
\label{czero}
\eeq
In Fig.~\ref{fig5}, we have also indicated the $C_0$ value as given by
(\ref{czero}) by dashed lines.  For the PDF close to the center of the flow this
estimate agrees with the value from the PDF of the data, but near the side
planes deviations become visible.  Here the strength of energy input is largest
on average and coherent flow structures are most prominent.  In all cases our
$C_0$ is larger than unity.  A $C_0$ larger than unity was also found in Farago's
model of Brownian particle motion where probabilities of time averaged energy
injection rates were considered \cite{Farago}.

\section{Concluding remarks}
We have studied the fluctuations of the energy injection rate in a turbulent
shear flow for various Reynolds numbers.  The mean value of the energy input
rate and the shape of its probability distribution varies across the shear flow.
The results do not depend much on the Reynolds number and the probability
density functions collapse when rescaled by the mean value of $\eps_{in}$.  The
probabilities for positive and negative energy injection are
exponentially related over a range of more than twice the mean values.  However,
the constant in the exponent varies across the layer.  Thus, while there are
some indications for an exponential relation between positive and negative
energy uptake, the quantitative behaviour depends on the position across the
layer.

It is still an open, but interesting, question whether the energy injection rate
(smoothed in time or not) can be considered as a {\em macroscopic} 
substitute for the entropy production rate, and whether fluctuation relations
for it can be found. As mentioned in the introduction, the Navier-Stokes
equation is perhaps closest in its global features to the Brownian particle
studied by Farago \cite{Farago}.  That model predicts a crossover to another
slope for large energy uptake, but the statistical uncertainties in our
numerical data are too large to study this.  Other systems that might provide
guidance for what to expect in the fluctuation statistics are thermostatted
Lorentz gases where time reversibility is broken by a magnetic field
\cite{Dolow02}.

Further avenues worth exploring include the relation between coherent structures
and energy uptake or energy blocking (as in Fig.~\ref{fig1}), and the
implications for backscatter effects in turbulent flows, i.e., the phenomenon
that locally energy does not cascade down to smaller but up to larger scales.
This might be important for subgrid-scale modelling within large-eddy
simulations \cite{Lesieur}.

\noindent
{\bf Acknowledgements}\\

We thank J.~Davoudi, J.~R.~Dorfman, G.~Gallavotti, W.~I.~Goldburg, W.~Losert, 
L.~Rondoni, T.~Tel and J.~Vollmer for useful discussions.  
The computations were done on a Cray
SV1ex at the John von Neumann-Institut f\"ur Computing at the Forschungszentrum
J\"ulich and we are grateful for their support.  This work was also supported by
the Deutsche Forschungsgemeinschaft.


\begin{thebibliography}{10}
\bibitem{Frisch} U.~Frisch, Turbulence,
                 Cambridge University Press, Cambridge 1995.
                 
\bibitem{Ciliberto98} S.~Ciliberto and C.~Laroche, J. Phys. IV France 8,
                      (1998) 215.

\bibitem{Goldburg} W.~I.~Goldburg, Y.~Y.~Goldschmidt, and H.~Kellay,
                   Phys. Rev. Lett. 87, (2001) 245502.

\bibitem{Pinton98} S.~T.~Bramwell, P.~C.~W.~Holdsworth and J.-F.~Pinton,
	Nature 396, (1998) 552.

\bibitem{Pinton99} J.-F.~Pinton, P.~C.~W.~Holdsworth and R.~Labb\'{e},
                   Phys.~Rev.~E 60, (1999) R2452.
	
\bibitem{Pinton2002} A. Noullez and J.-F.~Pinton, Eur. Phys. J. B 28, (2002)
	231.
                   
\bibitem{Cadot02} J.~H.~Titon and O.~Cadot, Phys. Fluids 15, (2003) 625.
                   
\bibitem{Evans93} D.~J.~Evans, E.~G.~D.~Cohen and G.~P.~Moriss,
                  Phys. Rev. Lett. 71, (1993) 2401.

\bibitem{GC95} G.~Gallavotti and E.~G.~D.~Cohen, Phys. Rev. Lett. 74, (1995)
               2694;  J. Stat. Phys. 80, (1995) 931.

\bibitem{Vollmer} J. Vollmer, Phys. Rep. 372, (2002) 131.

\bibitem{Farago} J. Farago, J. Stat. Phys. 107, (2002) 781.

\bibitem{She93} Z.-S. She and E. Jackson, Phys. Rev. Lett. 70, (1993) 1255.

\bibitem{Rondoni} L. Rondoni and E. Segre, Nonlinearity 12, (1999) 1471.

\bibitem{Gallavotti02} G. Gallavotti, L. Rondoni and E. Segre,
{\em Lyapunov spectra and nonequilibrium ensembles equivalence in 2D
fluid mechanics}, preprint (2002).

\bibitem{Aumaitre01} S. Auma\^itre, S. Fauve, S. McNamara and P. Poggi,
                     Eur. Phys. J. B 19, (2001) 449.
                     
\bibitem{Schu01} J.~Schumacher and B.~Eckhardt, Phys. Rev. E 63, (2001) 046307.

\bibitem{Landau} L. D. Landau and E. M. Lifschitz, {\em Course in
                 Theoretical Physics: Fluid Mechanics}, Pergamon Press,
                 Oxford, 1987.

\bibitem{Schu00} J.~Schumacher and B.~Eckhardt, Europhys. Lett.
	52, (2000) 627.

\bibitem{Dolow02} M.~Dolowschi\'{a}k and Z.~Kov\'{a}cs, 
Phys. Rev. E 66, (2002) 066217.

\bibitem{Lesieur} M.~Lesieur and O.~Metais, Annu. Rev. Fluid Mech. 28, (1996) 45. 


\end{thebibliography}
\end{document}